\documentclass[twocolumn,english,conference]{IEEEtran}
\usepackage[T1]{fontenc}
\usepackage[latin9]{inputenc}
\usepackage{geometry}
\usepackage{float}
\usepackage{units}
\usepackage{mathrsfs}
\usepackage{amsthm}
\usepackage{amsmath}
\usepackage{amssymb}
\usepackage{graphicx}

\makeatletter

\floatstyle{ruled}
\newfloat{algorithm}{tbp}{loa}
\providecommand{\algorithmname}{Algorithm}
\floatname{algorithm}{\protect\algorithmname}

\theoremstyle{plain}
\newtheorem{thm}{\protect\theoremname}

\usepackage{cite}

\IEEEoverridecommandlockouts

\makeatother

\usepackage{babel}
\providecommand{\theoremname}{Theorem}

\begin{document}

\title{Backhaul-Aware Caching Placement for Wireless Networks}

\author{Xi Peng$^{*}$, Juei-Chin Shen$^{\dagger}$, Jun Zhang$^{*}$ and
Khaled B. Letaief$^{*}$, \emph{Fellow}, IEEE\\
$^{*}$Department of ECE, The Hong Kong University of Science and
Technology\\
$^{\dagger}$Mediatek Inc., Hsinchu\\
E-mail: $^{*}$\{xpengab, eejzhang, eekhaled\}@ust.hk, $^{\dagger}$jc.shen@mediatek.com}
\maketitle
\begin{abstract}
As the capacity demand of mobile applications keeps increasing, the
backhaul network is becoming a bottleneck to support high quality
of experience (QoE) in next-generation wireless networks. Content
caching at base stations (BSs) is a promising approach to alleviate
the backhaul burden and reduce user-perceived latency. In this paper,
we consider a wireless caching network where all the BSs are connected
to a central controller via backhaul links. In such a network, users
can obtain the required data from candidate BSs if the data are pre-cached.
Otherwise, the user data need to be first retrieved from the central
controller to local BSs, which introduces extra delay over the backhaul.
In order to reduce the download delay, the caching placement strategy
needs to be optimized. We formulate such a design problem as the minimization
of the average download delay over user requests, subject to the caching
capacity constraint of each BS. Different from existing works, our
model takes BS cooperation in the radio access into consideration
and is fully aware of the propagation delay on the backhaul links.
The design problem is a mixed integer programming problem and is highly
complicated, and thus we relax the problem and propose a low-complexity
algorithm. Simulation results will show that the proposed algorithm
can effectively determine the near-optimal caching placement and provide
significant performance gains over conventional caching placement
strategies. \end{abstract}

\begin{IEEEkeywords}
Wireless caching networks, caching placement, download delay
\end{IEEEkeywords}

\IEEEpubid{}

\section{Introduction\label{sec:Introduction}}

The explosive growth of global mobile data traffic \cite{Cisco},
especially mobile video streaming, has led to significant increases
in user latency and imposed a heavy burden on backhaul links that
connect local BSs to the core network. The congestion in the backhaul
may cause excessively long delays to the content delivery, and thus
degrades the overall quality of experience (QoE). In order to support
the increasing mobile data traffic, one promising approach to reduce
delivery time and backhaul traffic is to deploy caches at BSs, thereby
bringing frequently requested bulky data (such as videos) close to
the users \cite{Golrezaei2013} in the hope of satisfying user requests
without increasing the burden over backhaul links. In this way, the
backhaul is used only to refresh the caches when the user request
distribution evolves over time. Usually, the refreshing process does
not require high-speed transmission and can work at off-peak times.

Caching capacity of local BSs can be regarded as a new type of resources
of wireless networks besides time, frequency and space. However, the
caching capacity is limited compared with the total amount of mobile
traffic. Thus sophisticated caching placement strategies will be needed
to fully exploit the benefit of caching. With the knowledge of channel
statistics and file popularity distribution, the central controller
is able to determine an optimal caching strategy to cater for user
requests with locally cached content. Once caches are fully utilized,
the requirements for backhaul can be greatly reduced and the download
delay can be shortened, especially when the backhaul links are in
poor conditions.

So far, the design of caching placement has not been fully addressed.
Most of the previous studies fail to take physical layer features
into consideration. For example, it was assumed in \cite{ProactiveStorage}
and \cite{FundamentalCaching} that the wireless transmission was
error-free. In \cite{D2D}, delays of D2D and cellular transmissions
were simply set as constants and the proposed caching strategy was
to store as many different files as possible. However, when taking
multipath fading into consideration, storing the same content at multiple
BSs can actually provide channel diversity gains and is perhaps more
advantageous. The authors in \cite{FemtoCaching} analyzed both uncoded
and coded femto-caching in order to minimize the total average delay
of all users. In their work, coded femto-caching was obtained as the
convex relaxation of the uncoded problem. Nevertheless, it not only
ignored physical layer features, but also imposed a certain network
topology requirement which cannot be fulfilled in practice. Physical-layer
operation including data assignment and coordinated beamforming in
caching networks was considered in \cite{OurPaper}, but the caching
placement was given as a prior.

There are also works studying the dynamic caching placement and update.
In \cite{videoAmazon}, the authors studied video caching in the radio
access network (RAN) and proposed caching policies based on the user
preference profile. Nevertheless, they considered neither the variation
of the wireless channel during the transmission of a file nor the
actual delay of wireless transmission and backhaul delivery. The authors
of \cite{CacheUpdate} concentrated on the caching content optimization
in a single BS. The file popularity was assumed unknown and their
strategy was optimized based on the observation of user request history
over time. However, this work did not consider the effect of backhaul
delays and assumed that the cache replacement was of negligible duration
and operated frequently.

In this paper, we present a wireless caching network model to determine
the optimal caching placement strategy for managing random user requests.
In particular, we aim at minimizing the average download delay, which
is one of the key QoE metrics, by taking wireless channel statistics
into account. Moreover, to the best of our knowledge, the impact of
backhaul delays on caching placement is studied in this paper for
the first time. The design of the caching placement strategy is formulated
as a mixed-integer nonlinear programming (MINLP) problem, which is
computationally difficult to solve. Thus we resort to the relaxed
formulation of the problem and provide a low-complexity algorithm.
Simulation results show that the strategy derived from our proposed
algorithm outperforms other well-known caching placement strategies.
Furthermore, we provide some insights into the caching placement design.
Specifically, in the case where the backhaul delay is very small,
the most popular content has a higher priority to be cached. On the
other hand, when the backhaul delay is relatively large, it should
be encouraged to maximize the \emph{caching content diversity}, i.e.,
to cache as much different content as possible, to reduce the chance
of invoking backhaul transmission.

\section{System Model }

\begin{figure}
\begin{centering}
\includegraphics[scale=0.8]{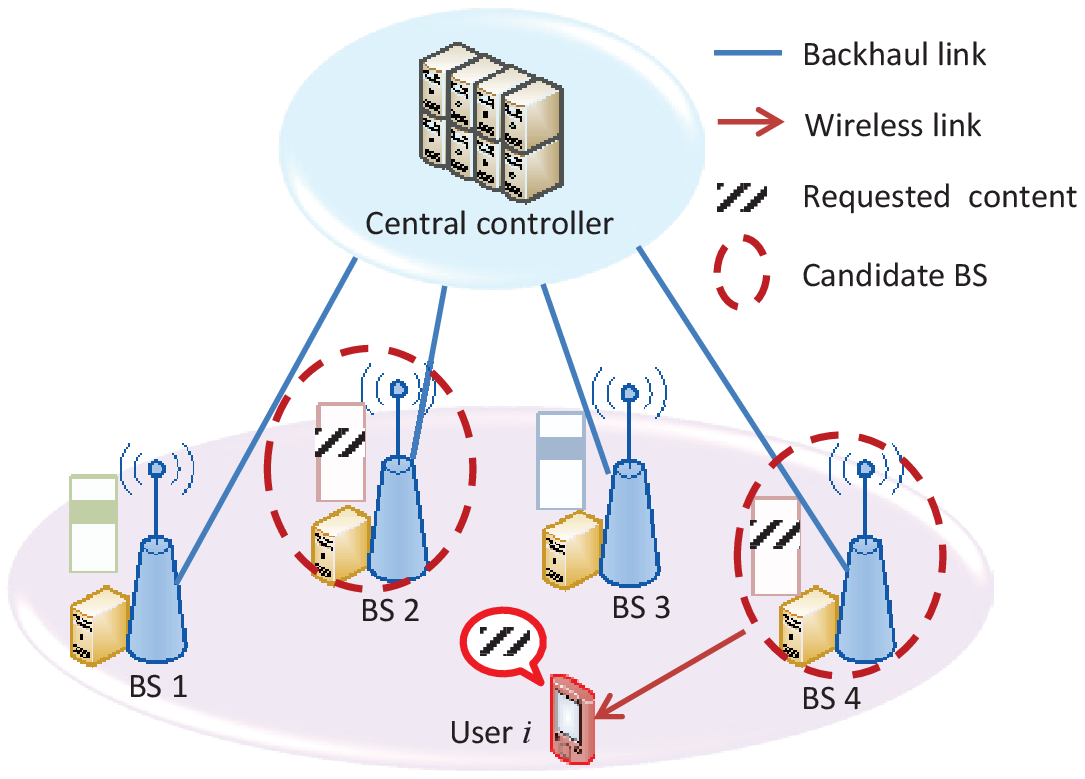}
\par\end{centering}

\begin{centering}
(a)
\par\end{centering}

\begin{centering}
\includegraphics[scale=0.8]{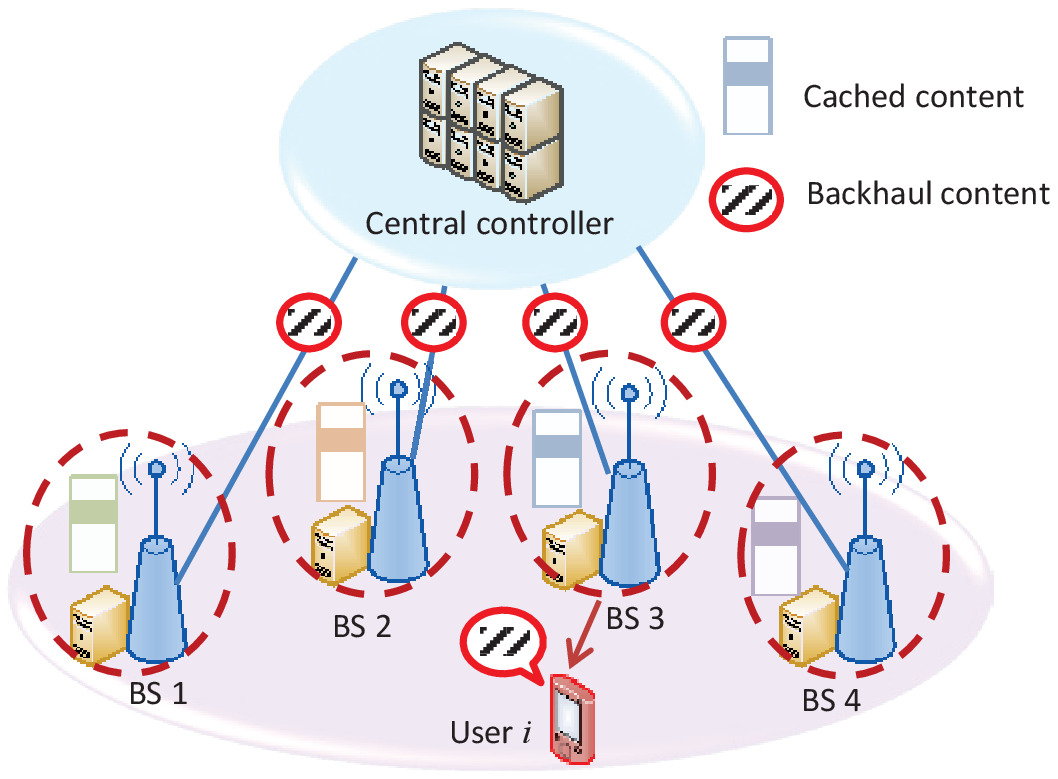}
\par\end{centering}

\begin{centering}
(b)
\par\end{centering}

\protect\caption{\label{fig:System-model-and}System model and content retrieval process.
(a) Content retrieval from only local BSs, $\Phi_{i}=\left\{ 2,4\right\} $;
(b) Content retrieval from the central controller with extra backhaul
delay, $\Phi_{i}=\left\{ 1,2,3,4\right\} $. }

\end{figure}

In this work, we consider the downlink of a wireless caching network,
in which $K$ single-antenna BSs and $U$ single-antenna mobile users
are uniformly distributed in the considered region. Through backhaul
links, the BSs are connected to a central controller, which also acts
as a file library server. The library contains $F$ files and each
of them can be divided into $L$ segments of equal size (with $b$
bits). Each BS is equipped with a storage unit of limited capacity.
For simplicity, we assume that all users see channels with the same
distribution. Without loss of generality, we focus on one user for
calculating the performance metric of interest. A BS will be regarded
as a candidate BS for user $i$ if it holds the content requested
by user $i$, no matter whether such content is previously cached
or retrieved via the backhaul. User $i$ has its cluster of candidate
BSs, denoted as $\Phi_{i}$, which consists of BS indices and has
cardinality $\left|\Phi_{i}\right|$. The system model and data retrieval
process are illustrated in Fig. \ref{fig:System-model-and}. The user
will choose the BS of the best wireless channel in $\Phi_{i}$ to
communicate with, as shown in Fig. \ref{fig:System-model-and} (a).
As for the case that no BS holds the requested content, as shown in
Fig. \ref{fig:System-model-and} (b), the central controller will
pass such content to all BSs and we shall have $\left|\Phi_{i}\right|=K$.

\subsection{File Transmission Model}

We consider segmented file transfer \cite{proxyCacheSeg} (also known
as multi-source file transfer), which has the advantage of allowing
a requested file to be sourced and downloaded from different BSs in
various time slots. Different segments of a requested file, when reaching
the user, will be decoded and assembled into a complete file. Each
segment can be as small as a single packet or as big as a stream lasting
for several minutes. We will first discuss how to calculate the delay
of downloading a segment, which is typically quantified as the average
number of required time slots for a segment to be successfully decoded.
Under the assumption that the segments of a file are subsequently
sent over homogenous and ergodic channels, the download delay of this
file is indeed the sum of all segment-download delays.

We shall assume that time is partitioned into $\tau$-second-wide
time slots indexed by $s$. We will also assume that at each time
slot, user $i$ communicates with the BS in $\Phi_{i}$ that provides
the highest signal-to-interference-plus-noise ratio (SINR), and this
BS is able to transmit $N\gg1$ complex symbols. Let $h_{i}\left[s\right]\in\mathbb{C}$
denote the channel coefficient between the selected BS and user $i$,
following the same distribution for all $i$ and $s$. The channel
is assumed to be block-fading with block length $\tau$, i.e., the
channel is constant for a duration of $\tau$ seconds and different
channel block realizations are i.i.d.. At time slot $s$, the transmit
signal from the selected BS to user $i$ is denoted as

\begin{equation}
\boldsymbol{x}_{i}\left[s\right]\triangleq\left(x_{i1}\left[s\right],\dots,x_{iN}\left[s\right]\right)\label{eq:Tx}
\end{equation}
and the transmit power constraint is $\mathbb{E}\left\{ \left|x_{in}\left[s\right]\right|^{2}\right\} \leq P_{t}$
for all $i\in\left\{ 1,\dots,U\right\} $, $n\in\left\{ 1,\dots,N\right\} $
and $s$.

The received signal of user $i$ for time slot $s$ can be written
as

\begin{equation}
\boldsymbol{y}_{i}\left[s\right]=h_{i}\left[s\right]\boldsymbol{x}_{i}\left[s\right]+\boldsymbol{n}_{i}\left[s\right],\label{eq:Rx}
\end{equation}
where $\boldsymbol{y}_{i}\left[s\right]\triangleq\left(y_{i1}\left[s\right],\dots,y_{iN}\left[s\right]\right)$
and $\boldsymbol{n}_{i}\left[s\right]\triangleq\left(n_{i1}\left[s\right],\dots,n_{iN}\left[s\right]\right)$.
The noise is complex white Gaussian, i.e., $\boldsymbol{n}_{i}\left[s\right]\thicksim\mathcal{C}\mathcal{N}\left(0,\sigma^{2}\right)$. 

The transmission of a segment to the intended user, in accordance
with an incremental redundancy (IR) hybrid-ARQ protocol as adopted
in LTE \cite{bookLTE}, will be presented below. Given a well-designed
Gaussian codebook, the $b$-bit file segment is encoded into a channel
code $\boldsymbol{c}=\left\{ \left.c_{1}\right|\left.c_{2}\right|\dots\left|c_{T}\right.\right\} $,
where $T$ is a given positive integer, which can be made arbitrarily
large, and each subcode $c_{j}$, containing $N$ complex symbols,
can be individually used to recover the original file segment. These
subcodes are further modulated into a series of signal bursts. At
each time slot, one signal burst is allowed to be sent. If this burst
is not decoded correctly, the user feeds back a negative acknowledgment
(NACK) message over an error-free and low-latency channel. Once the
BS receives this NACK message, the next burst is sent at the next
time slot. This process continues until the BS receives an acknowledgement
(ACK) message. If the transmission starts from the first time slot
and ends at the $S$-th time slot, the effective coding rate is $R/S$,
where the coding rate is defined as $R\triangleq b/N$ bits/sec/Hz
as a complex symbol can be transmitted approximately in 1 s and 1
Hz \cite{HARQ}.

\subsection{Caching Placement Strategy }

In practice, the caching capacity at BSs cannot be arbitrarily large.
Thus, it is crucial to optimize the caching strategy to exploit the
limited caching capacity in order to maximize the benefits brought
by caching. 

We denote the caching placement strategy as a three-dimensional matrix
$\mathcal{C}=\left[C_{1},\dots C_{k},\dots,C_{K}\right]\in\left\{ 0,1\right\} ^{F\times L\times K}$
consisting of binary entries, where $C_{k}$, indicating the caching
placement of BS $k$, is defined as  
\begin{equation}
C_{k}\triangleq\left[\begin{array}{ccc}
c_{11}^{k} & \dots & c_{1L}^{k}\\
\vdots & \ddots & \vdots\\
c_{F1}^{k} & \dots & c_{FL}^{k}
\end{array}\right]\in\left\{ 0,1\right\} ^{F\times L},\label{eq:C_k}
\end{equation}
in which $c_{fl}^{k}=1$ means that segment $l$ of file $f$ is cached
at BS $k$ in advance and $c_{fl}^{k}=0$ means the opposite. Since
each BS has a limited caching capacity $\bar{C}$, we have
\begin{equation}
\sum_{f=1}^{F}\sum_{l=1}^{L}c_{fl}^{k}\leq\bar{C},\quad\forall k\in\left\{ 1,\dots,K\right\} .\label{eq:Capacity}
\end{equation}

When user $i$ requests segment $l$ of file $f$, the number of candidate
BSs holding the segment is 
\begin{equation}
N_{fl}^{C}=\sum_{k=1}^{K}\mbox{c}_{fl}^{k}.\label{eq:N_fl}
\end{equation}
Note that there should be a constraint $0\leq N_{fl}^{C}\leq K$ for
$\forall f\in\left\{ 1,\dots,F\right\} ,\forall l\in\left\{ 1,\dots,L\right\} $
in order to avoid duplicate caching for the same segment at a BS.

\section{Backhaul-Aware Caching Placement}

The average download delay is a representative metric for the system
performance in wireless caching networks. In this section, we will
first derive an analytical expression of the download delay. We will
then formulate the problem of minimizing the download delay subject
to constraints on caching capacities.

\subsection{Download Delay}

We assume Rayleigh fading channels and no interference among different
users, e.g., they can be served by different subcarriers with orthogonal
frequency-division multiple access (OFDMA). Denote the signal-to-noise
ratio (SNR) from BS $k$ to user $i$ as $\textrm{ SNR}_{i}^{k}$,
and the probability density function (PDF) of $\textrm{SNR}_{i}^{k}$
is given by 
\begin{equation}
P\left(x\right)=\frac{1}{\bar{\rho}}\exp\left(-\frac{x}{\bar{\rho}}\right),\label{eq:PDF_snr}
\end{equation}
where $\bar{\rho}$ is the average received SNR and we have $\bar{\rho}=P_{t}/\sigma^{2}$.
According to our previous assumption, at time slot $s$, the received
SNR of user $i$ can be obtained as

\begin{equation}
\rho_{i}=\underset{k\in\Phi_{i}}{\max}\:\textrm{ SNR}_{i}^{k}.\label{eq:max_snr}
\end{equation}
Hence, the PDF of $\rho_{i}$ is
\begin{equation}
P_{\rho_{i}}\left(x\right)=\frac{\left|\Phi_{i}\right|}{\bar{\rho}}\exp\left(-\frac{x}{\bar{\rho}}\right)\left(1-\exp\left(-\frac{x}{\bar{\rho}}\right)\right)^{\left|\Phi_{i}\right|-1}.\label{eq:PDF_rho}
\end{equation}

For simplicity, we omit the subscript $i$ in the following derivation.

In the IR scheme, each user has a buffer with size $m$. Hence, up
to $m$ of the most recent signal bursts can be stored and used to
decode the information. In practice, $m$ is chosen to reach a compromise
between the decoding performance and the implementation cost. If the
buffer is big enough, $m$ can be regarded as infinity. If only the
current burst is used for decoding, we will have $m=1$. It has been
indicated in \cite{HARQ} that the mutual information across multiple
time slots can be written as 

\begin{equation}
R\left[s\right]=\sum_{j=\left(s-m\right)^{+}+1}^{s}\log_{2}\left(1+\rho\left[j\right]\right),\label{eq:Rate}
\end{equation}
where $\rho\left[j\right]$ is the received SNR at time slot $j$.
When employing typical set decoding, the probability of decoding error
for user $i$ at time slot $s$ can be expressed as
\begin{equation}
q\left[s\right]=\textrm{Pr}\left\{ R\left[s\right]<R\right\} .\label{eq:error_prob}
\end{equation}
It is difficult to obtain a closed-form expression for the average
download delay. Hence, we shall derive a lower bound. 
\begin{thm}
\label{thm:1}For the IR hybrid-ARQ protocol, the average download
delay of a segment in this system model is lower bounded by 
\[
\mathbb{E}\left\{ D^{\textrm{seg}}\right\} \geq\frac{1}{1-\left(1-\exp\left(-\frac{2^{R/m}-1}{\bar{\rho}}\right)\right)^{m\left|\Phi\right|}}.
\]
\end{thm}
\begin{IEEEproof}
For a given user, the probability that its download delay $D^{\textrm{seg}}$
of a segment ($b$ bits) is larger than $d$ time slots is given by
\begin{equation}
\begin{aligned}\textrm{Pr}\left\{ D^{\textrm{seg}}>d\right\}  & =\textrm{Pr}\left\{ R\left[s\right]<R,\textrm{ for }s=1,2,\dots,d\right\} \\
 & =q^{d}\left[s\right].
\end{aligned}
\label{eq:prob_delay}
\end{equation}
The expected download delay for a segment can be obtained as 
\begin{equation}
\begin{aligned}\mathbb{E}\left\{ D^{\textrm{seg}}\right\}  & =\sum_{d=0}^{+\infty}\textrm{Pr}\left\{ D^{\textrm{seg}}>d\right\} \\
 & =\frac{1}{1-q\left[s\right]}.
\end{aligned}
\label{eq:D_seg}
\end{equation}
From (\ref{eq:Rate}) and (\ref{eq:error_prob}), we can get 
\begin{equation}
\begin{aligned}q\left[s\right] & =\textrm{Pr}\left\{ \sum_{j=\left(s-m\right)^{+}+1}^{s}\log_{2}\left(1+\rho\left[j\right]\right)<R\right\} \end{aligned}
.\label{eq:}
\end{equation}
Note that if $\log_{2}\left(1+\rho\left[j\right]\right)<\frac{R}{m}$
for $\forall j\in\left\{ 1,\dots,s\right\} $, we have $\sum_{j=\left(s-m\right)^{+}+1}^{s}\log_{2}\left(1+\rho\right)<R$.
As a result, we can get 
\begin{equation}
\begin{aligned}q\left[s\right] & \geq\textrm{Pr}\left\{ \log_{2}\left(1+\rho\left[j\right]\right)<\frac{R}{m},\textrm{ for }\forall j\right\} \\
 & =\left(\textrm{Pr}\left\{ \log_{2}\left(1+\rho\right)<\frac{R}{m}\right\} \right)^{m}.
\end{aligned}
\label{eq:-1}
\end{equation}
With (\ref{eq:PDF_rho}), we can obtain the lower bound of the error
probability as 
\begin{equation}
\begin{aligned}q\left[s\right] & \geq\left(1-\exp\left(-\frac{2^{R/m}-1}{\bar{\rho}}\right)\right)^{m\left|\Phi\right|}\end{aligned}
.\label{eq:q_s_bound}
\end{equation}
Substituting (\ref{eq:q_s_bound}) into (\ref{eq:D_seg}) yields the
desired result.
\end{IEEEproof}
Theorem \ref{thm:1} implies the download delay for a segment is a
function of $\left|\Phi\right|$, $R$, $m$ and $\bar{\rho}$. We
will adopt the lower bound when calculating the average download delay
of a segment, which is given by 
\begin{equation}
D\left(\left|\Phi_{i}\right|,R,m,\bar{\rho}\right)=\frac{1}{1-\left(1-\exp\left(-\frac{2^{R/m}-1}{\bar{\rho}}\right)\right)^{m\left|\Phi_{i}\right|}}.\label{eq:D(Phi)}
\end{equation}

Note that if the requested segment has not been cached in any BS,
this segment will first be delivered to all $K$ BSs via backhaul
and then the wireless transmission will follow the same scheme as
the above-mentioned cached segment downloading. In that case, the
number of candidate BSs is $K$, i.e. , $\left|\Phi\right|=K$, and
an extra backhaul delay will be caused, which is denoted as $\delta$.

\subsection{Problem Formulation and Relaxation}

There are two cases when determining the candidate BSs. If $N_{fl}^{C}\neq0$,
we have $\left|\Phi\right|=N_{fl}^{C}$. Otherwise, when $N_{fl}^{C}=0$,
based on our assumption, we have $\left|\Phi\right|=K$ and the segment
download delay is $D\left(K,R,m,\bar{\rho}\right)+\delta.$ Accordingly,
the download delay of file $f$ can be calculated as
\begin{multline}
\sum_{l=1}^{L}\left\{ D\left(N_{fl}^{C},R,m,\bar{\rho}\right)\cdot\boldsymbol{1}\left(N_{fl}^{C}\neq0\right)\right.\\
+\left.\left[D\left(K,R,m,\bar{\rho}\right)+\delta\right]\cdot\boldsymbol{1}\left(N_{fl}^{C}=0\right)\right\} .\label{eq:delay_a_file}
\end{multline}
We suppose that the file $f$ is requested with probability $P_{f}$
and thus $\sum_{f=1}^{F}P_{f}=1$. The requests for segments of the
same file are of equal probability. Therefore, the average download
delay of all the files can be written as

\begin{gather}
f_{0}\left(\mathcal{C},R,m,\bar{\rho}\right)=\sum_{f=1}^{F}P_{f}\sum_{l=1}^{L}\left\{ D\left(N_{fl}^{C},R,m,\bar{\rho}\right)\cdot\boldsymbol{1}\left(N_{fl}^{C}\neq0\right)\right.\nonumber \\
\left.+\left[D\left(K,R,m,\bar{\rho}\right)+\delta\right]\cdot\boldsymbol{1}\left(N_{fl}^{C}=0\right)\right\} .\label{eq:delay_avg}
\end{gather}

Our goal is to minimize the average download delay by arranging the
placement of segments at each BS subject to the caching capacity limit,
given physical layer constraints, including the coding rate, the buffer
size at users and the received SNR target. With fixed $R$, $m$ and
$\bar{\rho}$, the caching placement problem is formulated as
\begin{align}
\mathscr{P}_{0}:\underset{\mathcal{C}}{\textrm{minimize}} & \quad f_{0}\left(\mathcal{C},R,m,\bar{\rho}\right)\label{eq:P0}\\
\textrm{subject to} & \quad\sum_{f=1}^{F}\sum_{l=1}^{L}c_{fl}^{k}\leq\bar{C},\forall k\in\left\{ 1,\dots,K\right\} \tag{C1}\nonumber \\
 & \quad0\leq\sum_{k=1}^{K}\mbox{c}_{fl}^{k}\leq K,\forall f,l\tag{C2}\nonumber \\
 & \quad\mathcal{C}\in\left\{ 0,1\right\} ^{F\times L\times K},\tag{C3}\nonumber 
\end{align}
where constraint C1 stands for the caching capacity limit of each
BS and constraint C2 indicates that each segment can be cached by
at most $K$ BSs. It turns out that problem $\mathscr{P}_{0}$ is
an MINLP problem and thus it is highly complicated to find the optimal
solution. As a result, we will focus on developing effective sub-optimal
algorithms.

By further examining problem $\mathscr{P}_{0}$, we find that the
optimization of caching placement boils down to the determination
of the number of candidate BSs for each segment. In order to simplify
the notations, we define it as a vector $\mathbf{x}=\left[x_{i}\right]\in\mathbb{N}^{FL}$,
where $x_{\left(f-1\right)F+l}=N_{fl}^{C}=\sum_{k=1}^{K}\mbox{c}_{fl}^{k}$.
We shall take two caching strategies as an example. That is, 
\begin{gather*}
\mathcal{C}_{1}=\left[\left[\begin{array}{cc}
1 & 0\\
1 & 0
\end{array}\right],\left[\begin{array}{cc}
1 & 1\\
0 & 0
\end{array}\right],\left[\begin{array}{cc}
1 & 0\\
1 & 0
\end{array}\right]\right],\\
\mathcal{C}_{2}=\left[\left[\begin{array}{cc}
1 & 0\\
1 & 0
\end{array}\right],\left[\begin{array}{cc}
1 & 0\\
1 & 0
\end{array}\right],\left[\begin{array}{cc}
1 & 1\\
0 & 0
\end{array}\right]\right].
\end{gather*}
If $\mathcal{C}_{1}$ is the optimal caching strategy, then $\mathcal{C}_{2}$
is also optimal. This is because both $\mathcal{C}_{1}$ and $\mathcal{C}_{2}$
correspond to the same vector $\mathbf{x}=\left[3,1,2,0\right]^{\mathrm{T}}$,
which determines the average download delay of the two strategies.
Therefore, (\ref{eq:D(Phi)}) can be written as 
\begin{equation}
D\left(x_{i}\right)=\frac{1}{1-\beta^{x_{i}}},\label{eq:Delay_seg_rewrt}
\end{equation}
with 
\begin{equation}
\beta\triangleq\left(1-\exp\left(-\frac{2^{R/m}-1}{\bar{\rho}}\right)\right)^{m}.\label{eq:beta}
\end{equation}
For $x_{i}>0$, we can find that $D\left(x_{i}\right)$ is convex
w.r.t. $x_{i}$ for $\forall i\in\left\{ 1,\dots,FL\right\} $. 

For each segment $x_{i},\textrm{ for }\forall i\in\left\{ 1,\dots,FL\right\} $,
its download delay is 
\begin{equation}
D\left(x_{i}\right)\cdot\boldsymbol{1}\left(x_{i}\neq0\right)+\left[D\left(K\right)+\delta\right]\cdot\boldsymbol{1}\left(x_{i}=0\right).\label{eq:D(xi)}
\end{equation}
The indicator functions in (\ref{eq:D(xi)}) will cause the major
difficulty in designing the caching placement strategy. To resolve
this issue, we adopt an exponential function $a^{x_{i}}$ with $0<a<1$
to approximate the indicator function $\boldsymbol{1}\left(x_{i}=0\right)$.
Then, we can obtain an approximated function $f\left(\mathbf{x}\right)$
to represent the average download delay of a file as 
\begin{alignat}{1}
f\left(\mathbf{x}\right) & =\sum_{i=1}^{FL}P_{\left\lceil \frac{i}{L}\right\rceil }\left\{ D\left(x_{i}\right)\cdot\left(1-a^{x_{i}}\right)+\left[D\left(K\right)+\delta\right]\cdot a^{x_{i}}\right\} \nonumber \\
 & =f_{1}\left(\mathbf{x}\right)+f_{2}\left(\mathbf{x}\right),\label{eq:f}
\end{alignat}
where $f_{1}\left(\mathbf{x}\right)$ and $f_{2}\left(\mathbf{x}\right)$
are given by

\begin{equation}
f_{1}\left(\mathbf{x}\right)=\sum_{i=1}^{FL}P_{\left\lceil \frac{i}{L}\right\rceil }\left\{ D\left(x_{i}\right)+\left[D\left(K\right)+\delta\right]\cdot a^{x_{i}}\right\} \label{eq:f1}
\end{equation}
and
\begin{equation}
f_{2}\left(\mathbf{x}\right)=-\sum_{i=1}^{FL}P_{\left\lceil \frac{i}{L}\right\rceil }D\left(x_{i}\right)a^{x_{i}}.\label{eq:f2}
\end{equation}
We can find that $f_{1}\left(\mathbf{x}\right)$ is convex w.r.t.
$\mathbf{x}$ and $f_{2}\left(\mathbf{x}\right)$ is concave w.r.t.
$\mathbf{x}$, which means that $f\left(\mathbf{x}\right)$ is the
difference of the convex (DC) functions.

With a fixed $\beta$, we consider an approximated problem $\mathscr{P}_{1}$
instead of $\mathscr{P}_{0}$. That is, 
\begin{align}
\mathscr{P}_{1}:\underset{\mathbf{x}}{\textrm{minimize}} & \quad f\left(\mathbf{x}\right)\label{eq:P1}\\
\textrm{subject to} & \quad\sum_{i=1}^{FL}x_{i}\leq K\bar{C}\nonumber \\
 & \quad0\leq x_{i}\leq K,\quad\forall i\in\left\{ 1,\dots,FL\right\} \nonumber \\
 & \quad\mathbf{x}\in\mathbb{N}^{FL}.\nonumber 
\end{align}
If we relax the integer constraint first, problem $\mathscr{P}_{1}$
turns out to be a DC programming problem, which is not easy to solve
directly due to the non-convex smooth objective function $f\left(\cdot\right)$.
The successive convex approximation (SCA) algorithm \cite{SCA} can
circumvent such a difficulty by replacing the non-convex object function
with a sequence of convex ones. Specifically, by starting from a feasible
point $\mathbf{x}^{\left(0\right)}$, the algorithm generates a sequence
$\left\{ \mathbf{x}^{\left(t\right)}\right\} $ according to the update
rule

\begin{equation}
\mathbf{x}^{\left(t+1\right)}=\mathbf{x}^{\left(t\right)}+\eta^{\left(t\right)}\left(\mathbf{\hat{x}}-\mathbf{x}^{\left(t\right)}\right),\label{eq:x_t+1}
\end{equation}
where $\mathbf{x}^{\left(t\right)}$ is the point generated by the
algorithm at the $t$-th iteration, $\eta^{\left(t\right)}$ is the
step size for the $t$-th iteration, and $\mathbf{\hat{x}}$ is the
solution of a convex optimization problem $\mathscr{Q}$, 
\begin{align}
\mathscr{Q}:\underset{\mathbf{\hat{x}}}{\textrm{minimize}} & \quad g\left(\mathbf{\hat{x}},\mathbf{x}^{\left(t\right)}\right)\label{eq:Q}\\
\textrm{subject to} & \quad\sum_{i=1}^{FL}\hat{x}_{i}\leq K\bar{C}\nonumber \\
 & \quad0\leq\hat{x}_{i}\leq K,\quad\forall i\in\left\{ 1,\dots,FL\right\} .\nonumber 
\end{align}
$g\left(\mathbf{x},\mathbf{x}^{\left(t\right)}\right)$ is an approximation
of $f\left(\mathbf{x}\right)$ at the $\left(t+1\right)$-th iteration,
which is defined as

\begin{equation}
g\left(\mathbf{x},\mathbf{x}^{\left(t\right)}\right)\triangleq f_{1}\left(\mathbf{x}\right)+\nabla f_{2}\left(\mathbf{x}^{\left(t\right)}\right)^{\mathrm{T}}\left(\mathbf{x}-\mathbf{x}^{\left(t\right)}\right)+\tau\left\Vert \mathbf{x}-\mathbf{x}^{\left(t\right)}\right\Vert _{2}^{2}\label{eq:g}
\end{equation}
and is a tight convex upper-bound of $f\left(\mathbf{x}\right)$.
The main steps of the algorithm are presented in Algorithm \ref{alg:}.
The solution obtained from Algorithm \ref{alg:} then should be rounded
due to the constraint $\mathbf{x}\in\mathbb{N}^{FL}$. 

\begin{algorithm}
\protect\caption{\textbf{:} \label{alg:} The SCA Algorithm}

\textbf{Initialization}: Find a feasible point $\mathbf{x}^{\left(0\right)}$
and set $t=0$.

\textbf{Repeat}

\qquad{}Solve problem $\mathscr{Q}$ and obtain the solution $\mathbf{\hat{x}}$;

\qquad{}Update $\mathbf{x}^{\left(t+1\right)}=\mathbf{x}^{\left(t\right)}+\eta^{\left(t\right)}\left(\mathbf{\hat{x}}-\mathbf{x}^{\left(t\right)}\right)$;

\qquad{}Set $t=t+1$;

\textbf{Until }stopping criterion is met.
\end{algorithm}

\section{Simulation Results}

In this section, we present numerical results to examine the performance
of the proposed caching placement strategy and to investigate the
impact of backhaul delay. 

Some previous studies have shown that in practical networks the request
probability of content can be fitted with some popularity distributions.
In this work, we assume that the popularity of files follows a Zipf
distribution with parameter $\gamma_{r}=0.6$ (see \cite{Youtube})
and the files are sorted in a descending order in terms of popularity.
We set the rate target as $R=2.5$ bits/sec/Hz and the average received
SNR as $\bar{\rho}=\unit[10]{dB}$. We consider the case where only
the current burst is used for decoding at the user side; i.e., $m=1$.
The range of the backhaul delivery delay $\delta$ is selected on
the basis of a measurement operated on a practical network, as was
done in \cite{videoAmazon}. Their experiment implied that the backhaul
delay of a piece of content approximately ranges from 30\% to 125\%
of its wireless transmit delay. To investigate the impact of such
delay, we choose $\delta\in\left[0,4\right]$.

\begin{figure}
\begin{centering}
\includegraphics[scale=0.6]{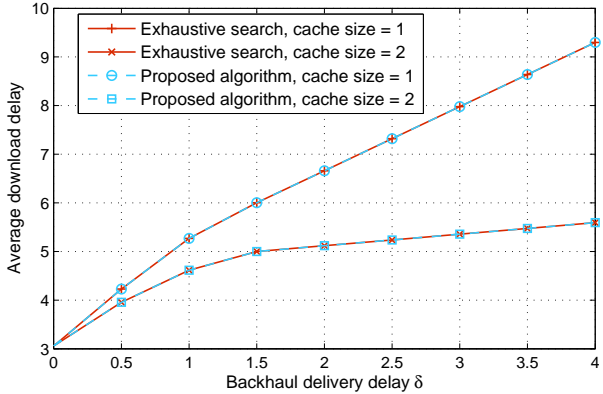}
\par\end{centering}

\protect\caption{\label{fig:Average-downloading-delay}Average download delay versus
backhaul delivery delay. }

\end{figure}

First we compare the performance of the proposed algorithm with exhaustive
search. The file library has three files and each of them is divided
into three segments. There are four BSs and each has a capacity of
two segments. We adopt a step size of $\eta=1$ and the stopping criterion
idistributions given by $\left.\left\Vert \mathbf{x}^{\left(t+1\right)}-\mathbf{x}^{\left(t\right)}\right\Vert _{2}\right/\left\Vert \mathbf{x}^{\left(t\right)}\right\Vert _{2}<10^{-4}.$
Fig. \ref{fig:Average-downloading-delay} shows that the results given
by the proposed efficient algorithm are very close to those obtained
using the exhaustive search. It can also be observed that a slight
increase in cache size $\bar{C}$ can significantly reduce the download
delay, which confirms that caching is of great potential to enhance
future wireless networks. 

\begin{figure}
\begin{centering}
\includegraphics[scale=0.6]{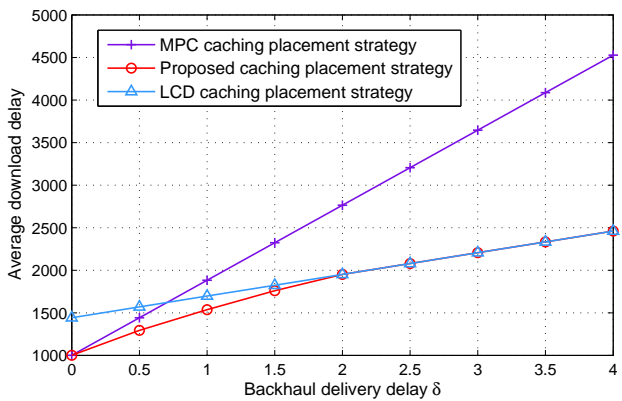}
\par\end{centering}

\protect\caption{\label{fig:Performance-comparison-of}Performance comparison of various
caching placement strategies.}

\end{figure}

Next, we compare the proposed algorithm with two standard caching
placement strategies on a large-scale system, where there are 50 BSs,
each with a capacity of $10^{4}$ segments. The file library has $10^{3}$
files, each of which is divided into $10^{3}$ segments. One standard
strategy stores $\bar{C}$ segments of the most popular content (MPC)
\cite{videoAmazon} in the cache memory of each BS. The other one
always places $K\bar{C}$ different segments in total at all BS caches,
which aims at the largest content diversity (LCD) \cite{D2D}. For
the MPC policy, $x_{1}=x_{2}=\cdots=x_{K}=\bar{C}$ and $x_{K+1}=\cdots=x_{FL}=0$;
and for the LCD policy, $x_{1}=x_{2}=\cdots=x_{K\bar{C}}=1$ and $x_{K\bar{C}+1}=\cdots=x_{FL}=0$.
The MPC policy is often adopted when multiple users access a few pieces
of content very frequently, such as popular movies and TV shows. The
LCD policy ensures that for most requests, part of the segments can
be served by the local caches, instead of having to be fetched from
the remote central controller through the backhaul. The step size
and the stopping criterion are the same as Fig. \ref{fig:Average-downloading-delay}.
Fig. \ref{fig:Performance-comparison-of} demonstrates that our proposed
strategy outperforms the other schemes, and important insights are
revealed. When the backhaul delay is small enough, it can be regarded
as equivalent to the case of infinite caching capacity. In this case,
the best way to save download time is to maximize the channel diversity
for each segment. As a result, the proposed strategy and the MPC policy
coincide at the point $\delta$=0. When the backhaul delay increases,
the advantage of caching diversity emerges, and the LCD policy will
surpass the MPC policy. Our proposed strategy and the LCD policy will
converge when backhaul links are suffering from severe delivery delays.
This is because when there is a large $\delta$, even a single delivery
via backhaul will lead to a huge delay and thus backhaul transmission
should be prevented as much as possible. Therefore, the strategy that
can provide maximum caching content diversity is favorable.

\section{Conclusions}

This paper presented a framework to minimize the average download
delay of wireless caching networks. A caching placement problem which
takes into account physical layer processing as well as backhaul delays
was formulated to fully exploit the benefit of caching. As the design
problem is an MINLP problem, we relaxed it into a DC optimization
problem and adopted the SCA algorithm to solve it efficiently. Simulation
results showed that our strategy can significantly reduce the average
download delay compared to conventional strategies, and the proposed
low-complexity algorithm can achieve comparable performance to exhaustive
search. Moreover, we demonstrated that the backhaul propagation delay
will greatly influence the caching placement. Specifically, when the
backhaul delay becomes very small or very large, our proposed strategy
will gradually evolve to the MPC and the LCD strategy, respectively.
In particular, for a practical value of the backhaul delay, the proposed
caching placement serves as the best strategy. Therefore, it can be
concluded that our work provides a promising model to formulate the
download delay for wireless caching networks, and important insights
are given for determining the optimal caching placement strategy under
different backhaul conditions.

\bibliographystyle{IEEEtran}
\bibliography{Caching_placement_str}

\end{document}